\documentclass[aps,pra,twocolumn,superscriptaddress]{revtex4-2}
\usepackage{amsmath}
\usepackage{amssymb}
\usepackage{graphicx}
\usepackage{hyperref}
\hypersetup{
    colorlinks=true,
    linkcolor=blue,
    citecolor=blue,
    urlcolor=blue
}

\begin{document}

\title{Theory of Symmetry-Protected Two-Photon Coherence}

\author{Xuanying Lai}
\affiliation{Elmore Family School of Electrical and Computer Engineering, Purdue University, West Lafayette, Indiana 47907, USA}
\affiliation{Department of Physics and Astronomy, Purdue University, West Lafayette, Indiana 47907, USA}

\author{Shengwang Du}
\email{dusw@purdue.edu}
\affiliation{Elmore Family School of Electrical and Computer Engineering, Purdue University, West Lafayette, Indiana 47907, USA}
\affiliation{Department of Physics and Astronomy, Purdue University, West Lafayette, Indiana 47907, USA}
\affiliation{Purdue Quantum Science and Engineering Institute, Purdue University, West Lafayette, Indiana 47907, USA}

\author{Yue Jiang}
\email{joj134@pitt.edu}
\affiliation{Department of Physics and Astronomy, University of Pittsburgh, Pittsburgh, Pennsylvania 15260, USA}

\date{\today}

\begin{abstract}
In a recent article [Phys. Rev. Lett. \textbf{133}, 033601 (2024)], the coherence time of degenerate entangled photon pairs (biphotons) generated via backward spontaneous four-wave mixing in a cold atomic ensemble was shown to be immune to optical loss and dephasing. This finding is crucial for practical applications in quantum information processing, quantum communication, and networking, where loss is inevitable. However, the underlying mechanism for this loss- and dephasing-insensitive biphoton coherence time was insufficiently studied in the previous article, as quantum noise was not taken into account. In this work, we employ the Heisenberg-Langevin approach to study this effect and provide a rigorous theoretical proof of the symmetry-protected biphoton coherence by taking quantum noise into consideration, as compared to the perturbation theory in the interaction picture.

\end{abstract}

\maketitle

\section{Introduction}
Entangled photon pairs with ultra-narrow bandwidth and long coherence time play a crucial role in quantum information processing \cite{alllightstorage,fockstatestorage,Dustorage}, distributed quantum sensor networks \cite{Liu2021,Kim2024}, distributed quantum computing \cite{oxforddc} and long distance quantum communication \cite{Craddock2024,Neumann2022}, and have therefore attracted considerable interest \cite{Sharypov2011,Zhao2014,Wang2022,Li2023}. In a traditional scheme, biphoton can be generated using spontaneous parametric down-conversion (SPDC) in a nonlinear crystal \cite{burnham1970observation,harris1967observation,shih1988new, kwiat1995new,hong1987measurement}, typically resulting in a biphoton bandwidth of the order of terahertz and a coherence time of picoseconds. The coherence time of biphoton can be extended close to 1 $\mu$s by placing the nonlinear crystal inside a high-finesse optical cavity \cite{liu2020sub}. Further increasing the photon pair coherence time in SPDC requires optimizing material properties, engineering phase-matching conditions, and incorporating high-finesse optical cavities with precise fabrication techniques, which remain challenging. On the other hand, using near-resonant biphoton generation via spontaneous four-wave mixing (SFWM) in a cold atomic ensemble \cite{PhysRevLett.100.183603, Du2008, Zhao2014}, long-coherence-time biphotons can be guaranteed by reducing the group velocity ($V_g$) of one photon using electromagnetically induced transparency (EIT) \cite{boller1991observation,harris1997electromagnetically,lukin2001controlling}. Narrowband biphoton generation with a coherence time of $13.4$ $\mu s$ has been demonstrated in the cold atomic system in free space \cite{Wang2022}.

In atomic systems, further increasing biphoton coherence time is primarily limited by the inability to achieve higher atomic density, as well as atomic loss and dephasing. In the normal spontaneous four-wave mixing (SFWM) scheme \cite{Du2008}, the coherence time is constrained by the exponentially decaying temporal biphoton waveform. This decay profile is due to photon pairs generated at different regions of the atomic ensemble experiencing varying levels of loss \cite{Du2008, PRLsymbiph}. However, a recent study on narrowband biphotons generated via a degenerate four-wave mixing scheme has shown that coherence time can be preserved by spatial-temporal symmetry even in the presence of significant atomic loss and dephasing \cite{PRLsymbiph}. These findings overcome one of the crucial limiting factors, loss, in achieving long-coherence-time photon pairs, opening up new possibilities for practical quantum applications, e.g., long-distance entanglement distribution, even in the presence of realistic challenges like loss and dephasing.

However, previous explanations relied on the simplified ``cartoon" model and a perturbative calculation in the interaction picture \cite{PRLsymbiph}. The role of quantum noise induced by loss and dephasing is not discussed. In this work, we employ the Heisenberg-Langevin approach to comprehensively study the degenerate backward SFWM process, where quantum noise has been rigorously incorporated. We establish a rigorous theoretical framework and provide a proof of the symmetry-protected two-photon coherence in the presence of loss and dephasing.


\section{Heisenberg Picture}\label{sec:Heisenberg}



\begin{figure}[h!]
    \centering    \includegraphics[width=1.0\linewidth]{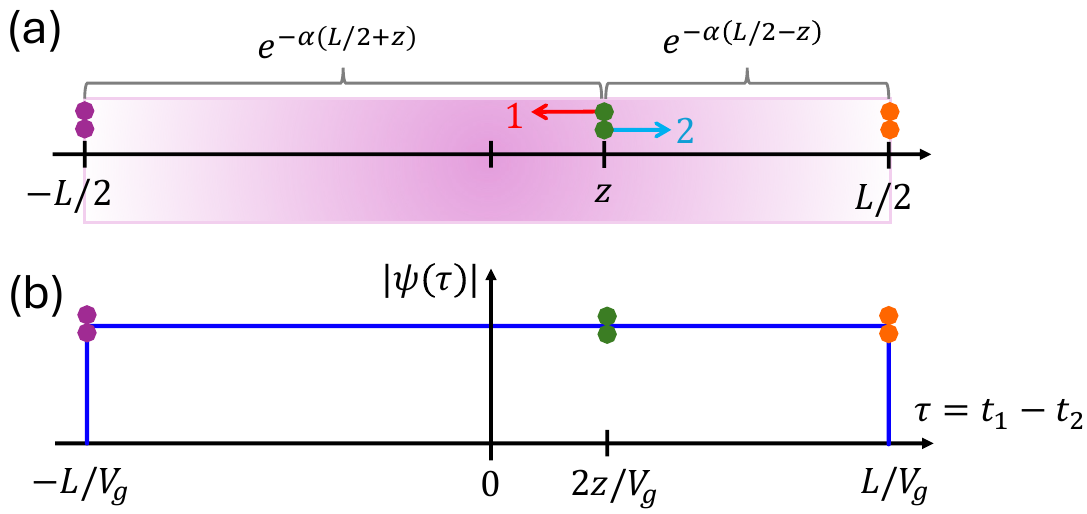}
    \caption{Cartoon picture of symmetry-protected two-photon coherence. (a) Backward-wave degenerate biphoton generation with an identical loss coefficient $\alpha$ for photons 1 and 2, propagating in opposite directions, leads to a two-photon joint probability (coincidence) amplitude attenuation $e^{-\alpha (L/2+z)}e^{-\alpha (L/2-z)}=e^{-\alpha L}$, which is independent of the photon-pair generation position $z$. (b) Biphoton coherence is quantified by measuring the two-photon coincidence amplitude as a function of the relative time delay $\tau = t_1 - t_2 = 2z / V_g$, corresponding to the arrival-time difference of photons 1 and 2 at the detectors. The coincidence amplitude exhibits a rectangular waveform with a coherence time of $2L / V_g$.}
    \label{fig:BiphotonGenerationSchematics}
\end{figure}

We consider backward-wave biphoton generation from a uniform nonlinear medium of length $L$, as shown in Fig.~\ref{fig:BiphotonGenerationSchematics}(a), with photon 1 propagating in the $-z$ direction and photon 2 in the $+z$ direction. Both photons experience the same absorption loss, characterized by the absorption coefficient $\alpha$, and slow group velocity $V_g\ll c$, where $c$ is the speed of light in vacuum. A photon pair can be generated at any point with uniform probability within the medium. Two single-photon counters are positioned at the two ends of the medium ($z=-L/2$ and $z=L/2$) to detect generated photon pairs. For paired photons generated at position $z$, the relative time difference for photon 1 arriving at $z=-L/2$ and photon 2 arriving at $z=L/2$ is $\tau=t_1-t_2= 2z/V_g$, registered as a coincidence count shown in Fig.~\ref{fig:BiphotonGenerationSchematics}(b), which displays a symmetric rectangular-shape two-photon joint probability amplitude function $\psi(\tau)$ with a coherence time determined by $2L/V_g$ \cite{PRLsymbiph}, which is twice as long as in the non-degenerate case (photon 1 and 2 have different frequencies) \cite{Du2008}. 

The positive-frequency parts of the generated fields are quantized as
\begin{equation}\label{eq:E}
\begin{aligned}
&E_1^{(+)}(z,t)=\sqrt{\frac{2\hbar\omega_0}{c\varepsilon_0A}}\hat a_1(z,t)e^{-i(\omega_0 t+k_0 z)},
\\ 
&E_2^{(+)}(z,t)=\sqrt{\frac{2\hbar\omega_0}{c\varepsilon_0A}}\hat a_2(z,t)e^{-i(\omega_0 t-k_0 z)}.
\end{aligned}
\end{equation}
with $k_0=\omega_0/c$ and $A$ being the single-mode cross-section area. Under the slowly varying envelop approximation, the two photon fields are governed by the following coupled Heisenberg-Langevin equations:
\begin{equation}\label{eq:HL01}
\begin{aligned}
i\begin{bmatrix}
\big(\frac{\partial}{\partial z}-\frac{1}{V_g}\frac{\partial}{\partial t}\big), 0\\
0,\big(\frac{\partial}{\partial z}+\frac{1}{V_g}\frac{\partial}{\partial t}\big)
\end{bmatrix}
\begin{bmatrix}
\hat{a}_1(z,t)\\
\hat{a}_2^\dagger(z,t)
\end{bmatrix}
&=
\mathbf{\hat H}_0
\begin{bmatrix}
\hat{a}_1 (z,t)\\
\hat{a}_2^\dagger(z,t)
\end{bmatrix} \\
&+i\sqrt{2\alpha}\begin{bmatrix}
\hat{f}_1(z,t) \\
\hat{f}_2^\dagger(z,t)
\end{bmatrix},
\end{aligned}
\end{equation}
where the effective Hamiltonian is 
\begin{equation}
\mathbf{\hat H}_0 =
\begin{bmatrix}
i\alpha &  {\kappa} \\
 {\kappa} & -i\alpha
\end{bmatrix},
\label{eq:H0}
\end{equation}
with $\kappa$ being the (real) nonlinear coupling coefficient, and $\hat f_{1,2}$ and $\hat f_{1,2}^{\dagger}$ being the Langevin noise (field) operators \cite{Jiang2023}, which are introduced due to loss. 

Interestingly, the operation on the left side of Eq.~\eqref{eq:HL01}
\begin{equation}
\begin{aligned}
&i\begin{bmatrix}
\big(\frac{\partial}{\partial z}-\frac{1}{V_g}\frac{\partial}{\partial t}\big), 0\\
0,\big(\frac{\partial}{\partial z}+\frac{1}{V_g}\frac{\partial}{\partial t}\big)
\end{bmatrix}
\end{aligned}
\end{equation}
is invariant under the following two transformations: (1) parity-inversion P: $z\rightarrow-z$, and (2) time reversal T: complex conjugation $i\rightarrow-i$, and $t\rightarrow-t$. The effective Hamiltonian $\mathbf{\hat H}_0$ in Eq.~(\ref{eq:H0}) also follows parity-time (PT) symmetry \cite{PhysRevLett.80.5243, Miri_2019, Jiang2023}. In conventional photonic systems, balanced gain and loss give rise to their PT symmetry \cite{Ruter2010,Lupu2013,Liu2016}. In the backward degenerate biphoton generation described in this work, the PT symmetry effectively ``turns" the loss of one mode into ``gain" to compensate for the loss in another mode so that the coherence of the two-photon joint amplitude is protected.



To derive the biphoton joint probability amplitude function, we take the following Fourier transform:
\begin{equation}
\begin{aligned}
\hat{a}_1(z,t)
&=\frac{1}{\sqrt{2\pi}}\int\hat{a}_1(z,\varpi)e^{-i\varpi t} d\varpi\\
\hat{a}_2(z,t)&=\frac{1}{\sqrt{2\pi}}\int\hat{a}_2(z,\varpi)e^{i\varpi t} d\varpi,
\end{aligned}
\label{eq:fields1}
\end{equation}
where we define $\hat{a}_1(z,\varpi)\equiv\hat{a}_1(z,\omega_0+\varpi)$ and $\hat{a}_2(z,\varpi)\equiv\hat{a}_2(z,\omega_0-\varpi)$, fulfilling energy conservation $\omega_1+\omega_2=2\omega_0$. Then Eq.~\eqref{eq:HL01} reduces to
\begin{equation}
\begin{aligned}
i\frac{\partial}{\partial z}
\begin{bmatrix}
\hat{a}_1(z,\varpi)\\
\hat{a}_2^\dagger(z,\varpi)
\end{bmatrix}
=
\mathbf{\hat H}
\begin{bmatrix}
\hat{a}_1 (z,\varpi)\\
\hat{a}_2^\dagger(z,\varpi)
\end{bmatrix} 
+i\sqrt{2\alpha}\begin{bmatrix}
\hat{f}_1(z,\varpi) \\
\hat{f}_2^\dagger(z,\varpi)
\end{bmatrix},
\end{aligned}
\label{eq:HL02}
\end{equation}
where the modified Hamiltonian is 
\begin{equation}
\mathbf{\hat H} =
\begin{bmatrix}
\frac{\varpi}{V_g}+i\alpha &  {\kappa} \\
 {\kappa} & -\frac{\varpi}{V_g}-i\alpha
\end{bmatrix},
\label{eq:H}
\end{equation}
and $\hat{f}_1(z,\varpi)$ and $\hat{f}_2^\dagger(z,\varpi)$ are the corresponding Fourier components of the Langevin field operators. 

With vacuum state input, the photon and Langevin field operators satisfy the following correlations: 
\begin{equation}
\begin{aligned}
&\left<\hat a_{m}^{\dagger}(z,\varpi_1),\hat a_{n}(z,\varpi_2)\right>=0,\\&\left<\hat a_{m}(z,\varpi_1),\hat a_{n}^{\dagger}(z,\varpi_2)\right>=\delta_{mn}\delta(\varpi_1-\varpi_2),\\
&\left<\hat a_{m}(z,\varpi_1),\hat a_{n}(z,\varpi_2)\right>=\left<\hat a_{m}^{\dagger}(z,\varpi_1),\hat a_{n}^{\dagger}(z,\varpi_2)\right>=0,\\
&\left<\hat f_{m}^\dag(z_1,\varpi_1)\hat f_{n}(z_2,\varpi_2)\right>=0,\\
&\left<\hat f_{m}(z_1,\varpi_1)\hat f_{n}^\dag(z_2,\varpi_2)\right>=\delta_{mn}\delta(\varpi_1-\varpi_2)\delta(z_1-z_2),\\
&\left<\hat f_{m}(z_1,\varpi_1)\hat f_{n}(z_2,\varpi_2)\right>=\left<\hat f_{m}^\dag(z_1,\varpi_1)\hat f_{n}^\dag(z_2,\varpi_2)\right>=0.
\end{aligned}\label{eq:general commutation}
\end{equation}

From Eq.~(\ref{eq:HL02}), we obtain solutions for $\hat a_1(-L/2)$ and $\hat a_2(L/2)$ \cite{Jiang2023} (here we omit the variable $\varpi$ for simplification):
\begin{widetext}
\begin{equation}
\begin{aligned}
\left[\begin{matrix}\hat a_{1}\left(-L/2\right)\\\hat a_2^\dag\left(L/2\right)\\\end{matrix}\right]
=\left[\begin{matrix}A&B\\C&D\\\end{matrix}\right]\left[\begin{matrix}\hat a_{1}\left(L/2\right)\\\hat a_2^\dag\left(-L/2\right)\\\end{matrix}\right]+\sqrt{2\alpha}\left[\begin{matrix}-A&0\\-C&1\\\end{matrix}\right]\int_{-L/2}^{L/2}\left[\begin{matrix}\bar A_1\left(z\right)&\bar B_1\left(z\right)\\\bar C_1\left(z\right)&\bar D_1\left(z\right)\\\end{matrix}\right]\left[\begin{matrix}\hat f_{1}\left(z\right)\\\hat f_2^\dag\left(z\right)\\\end{matrix}\right]dz,
\end{aligned}\label{eq:SolutionBW2}
\end{equation}
\end{widetext}
~\\
~\\
where 
$A=\frac{1}{\bar{A}(L)}$,
$B=-\frac{\bar{B}(L)}{\bar{A}(L)}$,
$C=\frac{\bar{C}(L)}{\bar{A}(L)}$,
$D=\bar{D}(L)-\frac{\bar{B}(L)\bar{C}(L)}{\bar{A}(L)}$, 
$e^{- i\mathbf{\hat H} L}\equiv\begin{bmatrix}
\bar{A} & \bar{B} \\
\bar{C} & \bar{D}
\end{bmatrix}$
and
$e^{- i\mathbf{\hat H} (L/2-z)}\equiv\begin{bmatrix}
\bar{A}_1(z) & \bar{B}_1(z) \\
\bar{C}_1(z) & \bar{D}_1(z)
\end{bmatrix}$.

The two-photon Glauber correlation can be computed by \cite{Glauber2006}
\begin{equation}\label{eq:g2}
\begin{aligned}
    G^{2}(\tau)&=\langle \hat{a}_2^{\dagger}(L/2,t_2)\hat{a}_1^{\dagger}(-L/2,t_1)\hat{a}_1(-L/2,t_1)\hat{a}_2(L/2,t_2)\rangle\\
    &=\langle \hat{a}_2^{\dagger}(L/2,t_2)\hat{a}_1^{\dagger}(-L/2,t_1)\rangle\langle \hat{a}_1(-L/2,t_1)\hat{a}_2(L/2,t_2)\rangle\\
    &+\langle \hat{a}_2^{\dagger}(L/2,t_2)\hat{a}_1(-L/2,t_1)\rangle\langle \hat{a}_1^{\dagger}(-L/2,t_1)\hat{a}_2(L/2,t_2)\rangle\\
    &+\langle \hat{a}_1^{\dagger}(-L/2,t_1)\hat{a}_1(-L/2,t_1)\rangle\langle \hat{a}_2^{\dagger}(L/2,t_2)\hat{a}_2(L/2,t_2)\rangle\\
    &=|\psi(\tau)|^2+|\psi'(\tau)|^2+R_1 R_2,
\end{aligned}
\end{equation}
where $\tau=t_1-t_2$. Here, we have applied the Gaussian moment theorem to decompose the fourth fields correlations to the sum of the products of second-order field correlations (see supplementary material of Ref. \cite{Jiang2023}). One can show that, with our solution in Eq.~\eqref{eq:SolutionBW2}, the second term in Eq.~\eqref{eq:g2} vanishes, i.e., $\psi'(\tau)=0$. $R_{1}$ and $R_{2}$ are the photon generation rates for field 1 and field 2, respectively, whose products contribute to the constant background accidental coincidence in $G^{(2)}(\tau)$. The biphoton temporal wave function, or the two-photon joint probability amplitude, is determined by
\begin{equation}\label{eq: heisenburg biphoton}
\begin{aligned}
   \psi(\tau)&= 
   \langle \hat{a}_1(-L/2,t_1)\hat{a}_2(L/2,t_2)\rangle\\
   &=\frac{1}{2\pi}\iint d\varpi_1 d\varpi_2 e^{-i\varpi_1 t_1}e^{i\varpi_2 t_2}\\
   &\quad\times\langle \hat{a}_1(-L/2, \varpi_1)\hat{a}_2(L/2, \varpi_2) \rangle\\
   &=\frac{1}{2\pi}\int d\varpi e^{-i\varpi\tau} \phi(\varpi),
\end{aligned}
\end{equation}
where 
\begin{widetext}
\begin{equation}\label{eq: result}
\begin{aligned}
    \phi(\varpi)
    =\underbrace{B(\varpi)D^{*}(\varpi)}_{\phi_0(\varpi)}\underbrace{-2\alpha A(\varpi)\int_{-L/2}^{L/2} \bar{B}_1(z,\varpi)\left[\bar{D}_1^*(z,\varpi)-C^* (\varpi)\bar{B}_1^*(z,\varpi)\right] dz}_{\phi_1(\varpi)}.
\end{aligned}
\end{equation}
\end{widetext}
The terms in Eq.~\eqref{eq: result} are 
\begin{equation}\label{eq: signal}
\begin{aligned}
    \phi_0(\varpi)= \frac{i \kappa \eta^* \sinh\left(L\eta\right)}{|\eta\cosh(L\eta)+\beta\sinh(L\eta)|^2},
\end{aligned}
\end{equation}
\begin{equation}\label{eq: Heisenburg noise}
\begin{aligned}
\phi_1(\varpi)
&=\frac{i2\kappa\alpha|\eta|^2\left[\cosh(L\eta)-\cosh\big(L\eta^*\,\big)\right]}{\left(\eta^{2}-\eta^{*2}\right)|\eta\cosh(L\eta)+\beta\sinh(L\eta)|^2}\\    &+\frac{i2\kappa\alpha\beta^{*}\left[\eta^*\sinh(L\eta)-\eta\sinh\big(L\eta^*\,\big)\right]}{\left(\eta^{2}-\eta^{*2}\right)|\eta\cosh(L\eta)+\beta\sinh(L\eta)|^2},
\end{aligned}
\end{equation}
with $\eta=\sqrt{\beta^2-\kappa^2}$ and $\beta=\alpha-i\varpi/V_g$. For the biphoton generation with small parameter gain, we take the approximation $\eta=\sqrt{\beta^2-\kappa^2}\simeq \beta$, which leads to
\begin{equation}\label{eq: psinew}
\begin{aligned}
\phi(\varpi)&\simeq\frac{i \kappa e^{-(\beta + \beta^*)L}(e^{\beta L} - e^{\beta^* L})}{\beta - \beta^*}
\\
&=i\kappa Le^{-\alpha L} \text{sinc}\left(\varpi L /V_g\right).
\end{aligned}
\end{equation}
The biphoton wavefunction in Eq.~\eqref{eq: heisenburg biphoton} becomes
\begin{equation}\label{eq:rectangular}
\begin{aligned}
   \psi(\tau) = \frac{i}{2} \kappa V_g \, e^{-\alpha L} \, \Pi\left(\tau; -L/V_g, L/V_g\right),
\end{aligned}
\end{equation}
where $\Pi$ is a unit rectangular function defined as $\Pi = 1$ for $\tau \in [-L/V_g, L/V_g]$ and $\Pi = 0$ otherwise. Thus we analytically derive the biphoton wavefunction from the Heisenberg-Langevin equations, providing a rigorous theoretical foundation for the results presented in Ref.~\cite{PRLsymbiph}. 

However, under the same approximation, without Langevin field, we have
\begin{equation}
\begin{aligned}\label{eq:BD*}
    \phi_0(\varpi)\simeq i\kappa Le^{-2\alpha L}\text{sinc}\left(\varpi L/V_g+i\alpha L\right),
\end{aligned}
\end{equation}
and its Fourier transformation is
\begin{equation}
\begin{aligned}\label{eq:BD*_time}
    \psi_0(\tau)= e^{-\alpha L}e^{-\alpha V_g \tau}\psi(\tau),
\end{aligned}
\end{equation}
which shows an exponentially decaying waveform with the presence of loss. Correspondingly, the Langevin field contributes to the wave function with
\begin{equation}
\begin{aligned}
    \psi_1(\tau)= (1-e^{-\alpha L}e^{-\alpha V_g \tau})\ \psi(\tau).
\end{aligned}
\end{equation}

Our analysis reveals that, within the Heisenberg-Langevin framework, the Langevin field operators play a critical role not only in preserving the commutation relations of the generated biphoton fields during propagation and evolution, but also in contributing significantly to the waveform of biphoton joint probability amplitude. Despite the presence of loss, which acts as a common attenuation factor on the entire waveform, the two-photon coherence time, determined by the relative group delay $2L/V_g$, is preserved.

\section{Interaction Picture}\label{sec:Interaction}

In the above Heisenberg picture, the state remains in the vacuum while the field operators evolve over time and space, and the biphoton wavefunction is computed from the field correlation in Eq.\ (\ref{eq: heisenburg biphoton}). In this section, we derive the biphoton state and wave function from the interaction picture, where the two counter-propagating fields 1 and 2 with symmetric loss are represented by their complex wave numbers $k_1=-(k_0+\varpi_1/V_g+i\alpha)$ and $k_2=k_0+\varpi_2/V_g+i\alpha$. The generated quantized fields illustrated in Fig.~\ref{fig:BiphotonGenerationSchematics} thus can be expressed as:
%
\begin{equation}\label{eq:E2}
\begin{aligned}
&\hat a_1(z,t)=\frac{e^{\alpha z}}{\sqrt{2\pi}}\int d\varpi_1 \hat a_1(\varpi_1
) e^{-i\varpi_1(t+z/V_g)},\\
&\hat a_2(z,t)=\frac{e^{-\alpha z}}{\sqrt{2\pi}}\int d\varpi_2 \hat a_2(\varpi_2) e^{-i\varpi_2(t-z/V_g)},
\end{aligned}
\end{equation}
where the annihilation operators in frequency domain satisfy the commutation relation $[\hat{a}_1(\varpi),\hat{a}^{\dagger}_1(\varpi')]=[\hat{a}_2(\varpi),\hat{a}^{\dagger}_2(\varpi')]=\delta(\varpi-\varpi')$. The interaction Hamiltonian for the SFWM process can be described as \cite{Du2008}
\begin{equation}
\begin{aligned}\label{eq:HI1}
\hat{H_I}
&=-\frac{c\varepsilon_0 A}{2\omega_0}\int_{-L/2}^{L/2} dz\ \kappa E_2^{(-)}(z,t)E_1^{(-)}(z,t)+H.c.\\
&=-\hbar \kappa \int_{-L/2}^{L/2} dz\ \hat{a}_2^\dagger(z,t) \hat{a}_1^\dagger(z,t)+H.c.
\end{aligned}
\end{equation} 
As shown in Eq. (\ref{eq:E2}), the symmetric loss induced factors $e^{\alpha z}$ and $e^{-\alpha z}$ cancel in the product $\hat{a}_2^\dagger(z,t) \hat{a}_1^\dagger(z,t)$. As a result, the interaction Hamiltonian becomes loss-independent:
\begin{equation}\label{eq:HI2}
\begin{aligned}
\hat{H_I}
&=-\frac{\hbar\kappa L}{2\pi} \iint  d\varpi_1 d\varpi_2 e^{i(\varpi_1+\varpi_2)t} \text{sinc}\left[\frac{(\varpi_{1}-\varpi_{2})L}{2V_g}\right] \\
&\times\hat{a}_2^\dagger(\varpi_2)\hat{a}_1^\dagger(\varpi_1)
\end{aligned}
\end{equation} 
Using first-order perturbation theory, we obtain the two-photon state 
\begin{equation}
\begin{aligned}
&|\Psi\rangle=-\frac{i}{\hbar}\int^{+\infty}_{-\infty} dt \hat{H_I}(t)|0\rangle\\
&=i\kappa L \iint d\varpi_{1}d\varpi_{2}\  \text{sinc}\left[\frac{(\varpi_{1}-\varpi_{2})L}{2V_g}\right]\\
&\times\delta(\varpi_1+\varpi_2) \hat{a}^{\dagger}_{2}(\varpi_{2})\hat{a}^{\dagger}_{1}(\varpi_{1})|0\rangle\\
&=i\kappa L\int d\varpi\ \text{sinc}\left(\varpi L/V_g\right)\hat{a}^{\dagger}_{2}(-\varpi)\hat{a}^{\dagger}_{1}(\varpi)|0\rangle,
\label{7}
\end{aligned}
\end{equation}
where the time integration results in Dirac $\delta$ function,  $\int e^{i(\varpi_1+\varpi_2)t} dt=2\pi\delta(\varpi_1+\varpi_2)$, indicating energy conservation $\varpi_1=-\varpi_2=\varpi$. 

The two-photon wavefunction, defined as the biphoton joint probability amplitude \cite{Du2008} is
\begin{equation}
\begin{aligned} 
\psi(\tau)&=\langle 0|\hat{a}_{1}(t_1,-L/2)\hat{a}_{2}(t_2,L/2)|\Psi\rangle\\
&=\frac{i\kappa L}{2\pi}e^{-\alpha L}\int d\varpi\ \text{sinc}\left(\varpi L/V_g\right)e^{-i 
\varpi \tau},\\
&=\frac{i}{2} \kappa V_g e^{-\alpha L} \ \Pi\left(\tau;-L/V_g,L/V_g\right),
\label{8}
\end{aligned}
\end{equation}
which is the same result as Eq.\ \eqref{eq:rectangular} obtained in the Heisenberg picture.

\section{Comparison of biphoton wavefunction in two pictures}\label{sec:5}
In both the Heisenberg and interaction pictures, the rectangular-shape biphoton waveform derived in Eqs.\ (\ref{eq:rectangular}) and (\ref{8}) relies on the small-parametric gain approximation $\kappa\ll|\beta|$. This assumption can break down for large nonlinear coupling $\kappa$. With large $\kappa$, the perturbative result from the interaction picture to the first order is no longer an accurate description for photon pair generation, and the Heisenberg picture method is preferred. To assess the impact of varying $\kappa$, we perform Heisenberg picture simulations without approximating $\eta = \sqrt{\beta^2 - \kappa^2}$ and plot the results based on Eq.\ (\ref{eq: result}), comparing them with those obtained using the interaction picture formalism (Eq.\ (\ref{8})).

\begin{figure}[h!]
    \centering
    \includegraphics[width=\linewidth]{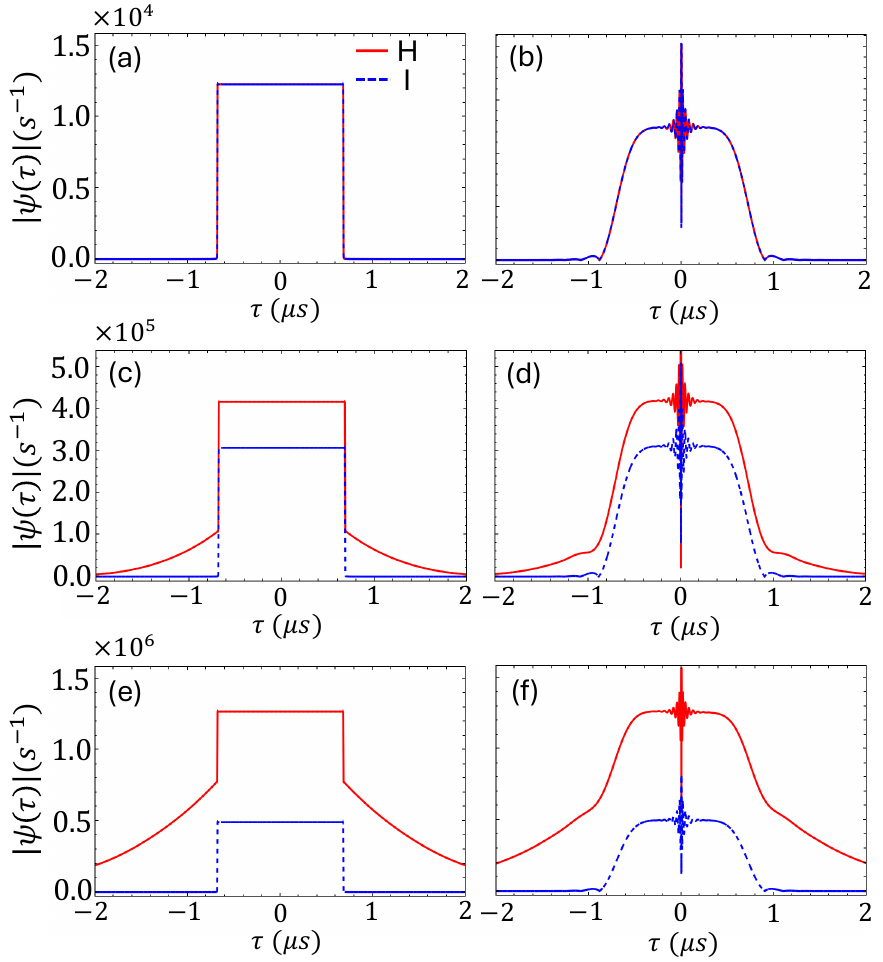}
    \caption{Numerical simulations of degenerate biphoton coincidence in interaction picture (I)  and Heisenberg picture (H). Common parameters: loss $\alpha(\varpi=0)L=0.51$ and group velocity $V_g(\varpi=0)=2.4\times10^4\ \text{m/s}$. The left column (a, c, e) shows simulations with constant $\kappa$, $\alpha$ and $V_g$, while the right column (b, d, f) corresponds to simulations based on realistic experimental conditions \cite{PRLsymbiph} with frequency-dependent $\kappa(\varpi)$, $\alpha(\varpi)$ and $V_g(\varpi)$. Nonlinear coupling is varied, (a) and (b): $\kappa(\varpi=0)L=0.03$. (c) and (d): $\kappa(\varpi=0)L=0.87$. (e) and (f): $\kappa(\varpi=0)L=1.40$.}
    \label{fig:comparison}
\end{figure}

Under the condition $\kappa\ll\alpha$, and $\alpha L=\alpha(\varpi=0)L=0.51$ and $V_g=V_g(\varpi=0)=2.4\times10^4\ \text{m/s}$, both pictures yield a rectangular biphoton waveform, as shown in Fig. \ref{fig:comparison}(a), consistent with the analytic results from Secs.\ \ref{sec:Heisenberg} and \ref{sec:Interaction}. As $\kappa$ increases, the Heisenberg picture shows clear deviations from the rectangular waveform. However, the interaction picture always produces a rectangular shape, even when it no longer accurately represents the true waveform—this discrepancy is visible in Fig. \ref{fig:comparison}(c) and (e).

In realistic experimental systems, the parameters $\kappa(\varpi)$, $\alpha(\varpi)$ and $V_g(\varpi)$ vary with frequency. We numerically simulate biphoton generation in a cold $^{87}\text{Rb}$ atomic ensemble \cite{PRLsymbiph}, and observe that the interaction picture with perturbation increasingly deviates from the actual waveform as $\kappa$ grows, while the Heisenberg picture continues to capture the correct dynamics. This behavior is illustrated in Fig. \ref{fig:comparison}(b), (d) and (f).



\begin{figure}[h!]
    \centering    \includegraphics[width=\linewidth]{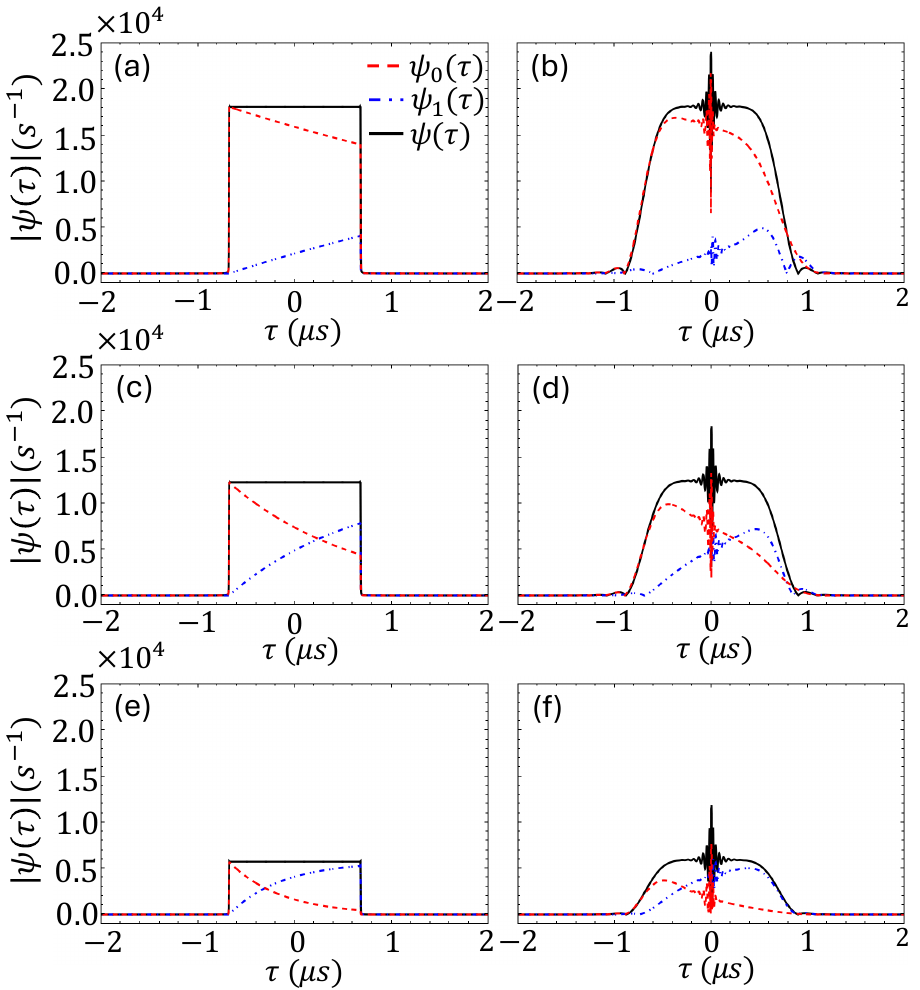}
    \caption{Impact of Langevin term on the biphoton waveform in the Heisenberg picture. Common parameters: nonlinear coupling $\kappa(\varpi=0)L=0.03$ and group velocity $V_g(\varpi=0)=2.4\times10^4\ \text{m/s}$. The left column (a, c, e) shows simulations with constant $\kappa$, $\alpha$ and $V_g$, while the right column (b, d, f) corresponds to simulations based on realistic SFWM experimental conditions \cite{PRLsymbiph} with frequency-dependent $\kappa(\varpi)$, $\alpha(\varpi)$ and $V_g(\varpi)$. Loss is varied, (a) and (b): $\alpha(\varpi=0)L=0.13$. (c) and (d): $\alpha(\varpi=0)L=0.51$. (e) and (f): $\alpha(\varpi=0)L=1.26$.}   \label{fig:signalandnoise}
\end{figure}

\section{Role of Langevin fields}
In the Heisenberg picture, as shown in Sec.~\ref{sec:Heisenberg}, the Langevin field operators play a critical role in the waveform of biphoton joint probability amplitude. If the contribution from Langevin field operators is not taken into account, the biphoton wavefunction $\psi_0(\tau)$ exhibits exponential decay due to system loss, as shown in Eq.\ (\ref{eq:BD*_time}). The Langevin field compensates for this decay and recovers the two-photon coherence, as revealed by Eq.\ (\ref{eq:rectangular}). In this section, we perform numerical simulations to illustrate the critical role of the Langevin field in preserving the coherence time by varying the loss $\alpha$.

Similarly to Sec.~\ref{sec:5}, we simulate the following two cases: 1) $\kappa$, $\alpha$ and $V_g$ are constants, and 2) they are frequency dependent in realistic SFWM experimental conditions \cite{PRLsymbiph}. As shown in Fig. \ref{fig:signalandnoise}, for both cases, $\psi_0(\tau)$ -- biphoton wavefunction without taking Langevin field into consideration, shows reduced coherence time as loss increases (red dashed lines). However, this effect is compensated for by $\psi_1(\tau)$ -- Langevin field contribution to the biphoton wavefunction (blue dashed lines). Therefore, the ovreall coherence in biphoton wavefunction $\psi(\tau)$ is preserved, as depicted by black solid lines.

Regardless of the loss, the Langevin field consistently compensates for the loss-induced decoherence, acting as a “restoring force” that counteracts the effects of loss and preserves the symmetry and correlation properties of the generated photon pairs. However, this compensation comes at the cost of a reduced biphoton generation rate and signal-to-noise ratio, quantified by $|\psi(\tau)|^2/(R_1R_2)$, where $R_{1}$ and $R_{2}$ denote the photon generation rates of photon 1 and photon 2, respectively.

\section{Conclusion}\label{sec:6}

In summary, we have established a rigorous theoretical framework for symmetry-protected coherence in degenerate backward biphoton generation, analyzed in both the Heisenberg and interaction pictures. In the regime of small parametric coupling $\kappa$, both approaches yield the same analytical expression for a rectangular-shaped biphoton temporal wavefunction, whose coherence time, $2L/V_g$, is preserved against loss and dephasing by space-time symmetry, and depends only on the length of the medium and the group velocity. The Langevin field plays a significant role in biphoton coherence time preservation, which acts as a “restoring force” that counteracts the effects of loss and preserves the waveform of the generated photon pairs. 

In the regime of strong parametric coupling, the Heisenberg-Langevin formalism remains valid and captures the effects of multiphoton processes, which lead to deviations from the ideal rectangular waveform. In contrast, the first-order perturbation theory in the interaction picture becomes inadequate for describing the two-photon state, as it neglects higher-order contributions arising from multiple photon-pair generation. Accurate modeling in this regime requires incorporating these higher-order corrections. 

In the backward degenerate biphoton generation, Parity-Time symmetry is inherent in the system Hamiltonian (Eq.\ (\ref{eq:H0})), due to the backward propagation of field 1, which effectively turns the ``loss" into ``gain" to compensate for the loss in field 2. However, the underlying physics explaining how this symmetry may result in a coherence-protected biphoton wave function has yet to be fully explored.

\begin{acknowledgments}
X.L and S.D. acknowledge the support from NSF (Nos. 2500662 and
2228725), Y.J acknowledges the support from Pittsburgh Quantum Institute community collaboration award. 
\end{acknowledgments}

\bibliography{main}

\begin{thebibliography}{31}%
\makeatletter
\providecommand \@ifxundefined [1]{%
 \@ifx{#1\undefined}
}%
\providecommand \@ifnum [1]{%
 \ifnum #1\expandafter \@firstoftwo
 \else \expandafter \@secondoftwo
 \fi
}%
\providecommand \@ifx [1]{%
 \ifx #1\expandafter \@firstoftwo
 \else \expandafter \@secondoftwo
 \fi
}%
\providecommand \natexlab [1]{#1}%
\providecommand \enquote  [1]{``#1''}%
\providecommand \bibnamefont  [1]{#1}%
\providecommand \bibfnamefont [1]{#1}%
\providecommand \citenamefont [1]{#1}%
\providecommand \href@noop [0]{\@secondoftwo}%
\providecommand \href [0]{\begingroup \@sanitize@url \@href}%
\providecommand \@href[1]{\@@startlink{#1}\@@href}%
\providecommand \@@href[1]{\endgroup#1\@@endlink}%
\providecommand \@sanitize@url [0]{\catcode `\\12\catcode `\$12\catcode `\&12\catcode `\#12\catcode `\^12\catcode `\_12\catcode `\%12\relax}%
\providecommand \@@startlink[1]{}%
\providecommand \@@endlink[0]{}%
\providecommand \url  [0]{\begingroup\@sanitize@url \@url }%
\providecommand \@url [1]{\endgroup\@href {#1}{\urlprefix }}%
\providecommand \urlprefix  [0]{URL }%
\providecommand \Eprint [0]{\href }%
\providecommand \doibase [0]{https://doi.org/}%
\providecommand \selectlanguage [0]{\@gobble}%
\providecommand \bibinfo  [0]{\@secondoftwo}%
\providecommand \bibfield  [0]{\@secondoftwo}%
\providecommand \translation [1]{[#1]}%
\providecommand \BibitemOpen [0]{}%
\providecommand \bibitemStop [0]{}%
\providecommand \bibitemNoStop [0]{.\EOS\space}%
\providecommand \EOS [0]{\spacefactor3000\relax}%
\providecommand \BibitemShut  [1]{\csname bibitem#1\endcsname}%
\let\auto@bib@innerbib\@empty
\bibitem [{\citenamefont {Hashimoto}\ \emph {et~al.}(2019)\citenamefont {Hashimoto}, \citenamefont {Toyama}, \citenamefont {Yoshikawa}, \citenamefont {Makino}, \citenamefont {Okamoto}, \citenamefont {Sakakibara}, \citenamefont {Takeda}, \citenamefont {van Loock},\ and\ \citenamefont {Furusawa}}]{alllightstorage}%
  \BibitemOpen
  \bibfield  {author} {\bibinfo {author} {\bibfnamefont {Y.}~\bibnamefont {Hashimoto}}, \bibinfo {author} {\bibfnamefont {T.}~\bibnamefont {Toyama}}, \bibinfo {author} {\bibfnamefont {J.-i.}\ \bibnamefont {Yoshikawa}}, \bibinfo {author} {\bibfnamefont {K.}~\bibnamefont {Makino}}, \bibinfo {author} {\bibfnamefont {F.}~\bibnamefont {Okamoto}}, \bibinfo {author} {\bibfnamefont {R.}~\bibnamefont {Sakakibara}}, \bibinfo {author} {\bibfnamefont {S.}~\bibnamefont {Takeda}}, \bibinfo {author} {\bibfnamefont {P.}~\bibnamefont {van Loock}},\ and\ \bibinfo {author} {\bibfnamefont {A.}~\bibnamefont {Furusawa}},\ }\bibfield  {title} {\bibinfo {title} {All-optical storage of phase-sensitive quantum states of light},\ }\href {https://doi.org/10.1103/PhysRevLett.123.113603} {\bibfield  {journal} {\bibinfo  {journal} {Phys. Rev. Lett.}\ }\textbf {\bibinfo {volume} {123}},\ \bibinfo {pages} {113603} (\bibinfo {year} {2019})}\BibitemShut {NoStop}%
\bibitem [{\citenamefont {Bouillard}\ \emph {et~al.}(2019)\citenamefont {Bouillard}, \citenamefont {Boucher}, \citenamefont {Ferrer~Ortas}, \citenamefont {Pointard},\ and\ \citenamefont {Tualle-Brouri}}]{fockstatestorage}%
  \BibitemOpen
  \bibfield  {author} {\bibinfo {author} {\bibfnamefont {M.}~\bibnamefont {Bouillard}}, \bibinfo {author} {\bibfnamefont {G.}~\bibnamefont {Boucher}}, \bibinfo {author} {\bibfnamefont {J.}~\bibnamefont {Ferrer~Ortas}}, \bibinfo {author} {\bibfnamefont {B.}~\bibnamefont {Pointard}},\ and\ \bibinfo {author} {\bibfnamefont {R.}~\bibnamefont {Tualle-Brouri}},\ }\bibfield  {title} {\bibinfo {title} {Quantum storage of single-photon and two-photon fock states with an all-optical quantum memory},\ }\href {https://doi.org/10.1103/PhysRevLett.122.210501} {\bibfield  {journal} {\bibinfo  {journal} {Phys. Rev. Lett.}\ }\textbf {\bibinfo {volume} {122}},\ \bibinfo {pages} {210501} (\bibinfo {year} {2019})}\BibitemShut {NoStop}%
\bibitem [{\citenamefont {Wang}\ \emph {et~al.}(2019)\citenamefont {Wang}, \citenamefont {Li}, \citenamefont {Zhang}, \citenamefont {Su}, \citenamefont {Zhou}, \citenamefont {Liao}, \citenamefont {Du}, \citenamefont {Yan},\ and\ \citenamefont {Zhu}}]{Dustorage}%
  \BibitemOpen
  \bibfield  {author} {\bibinfo {author} {\bibfnamefont {Y.}~\bibnamefont {Wang}}, \bibinfo {author} {\bibfnamefont {J.}~\bibnamefont {Li}}, \bibinfo {author} {\bibfnamefont {S.}~\bibnamefont {Zhang}}, \bibinfo {author} {\bibfnamefont {K.}~\bibnamefont {Su}}, \bibinfo {author} {\bibfnamefont {Y.}~\bibnamefont {Zhou}}, \bibinfo {author} {\bibfnamefont {K.}~\bibnamefont {Liao}}, \bibinfo {author} {\bibfnamefont {S.}~\bibnamefont {Du}}, \bibinfo {author} {\bibfnamefont {H.}~\bibnamefont {Yan}},\ and\ \bibinfo {author} {\bibfnamefont {S.-L.}\ \bibnamefont {Zhu}},\ }\bibfield  {title} {\bibinfo {title} {Efficient quantum memory for single-photon polarization qubits},\ }\href {https://www.nature.com/articles/s41566-019-0368-8} {\bibfield  {journal} {\bibinfo  {journal} {Nature photonics}\ }\textbf {\bibinfo {volume} {13}},\ \bibinfo {pages} {346} (\bibinfo {year} {2019})}\BibitemShut {NoStop}%
\bibitem [{\citenamefont {Liu}\ \emph {et~al.}(2021)\citenamefont {Liu}, \citenamefont {Zhang}, \citenamefont {Li}, \citenamefont {Zhang}, \citenamefont {Yin}, \citenamefont {Fei}, \citenamefont {Li}, \citenamefont {Liu}, \citenamefont {Xu}, \citenamefont {Chen},\ and\ \citenamefont {Pan}}]{Liu2021}%
  \BibitemOpen
  \bibfield  {author} {\bibinfo {author} {\bibfnamefont {L.-Z.}\ \bibnamefont {Liu}}, \bibinfo {author} {\bibfnamefont {Y.-Z.}\ \bibnamefont {Zhang}}, \bibinfo {author} {\bibfnamefont {Z.-D.}\ \bibnamefont {Li}}, \bibinfo {author} {\bibfnamefont {R.}~\bibnamefont {Zhang}}, \bibinfo {author} {\bibfnamefont {X.-F.}\ \bibnamefont {Yin}}, \bibinfo {author} {\bibfnamefont {Y.-Y.}\ \bibnamefont {Fei}}, \bibinfo {author} {\bibfnamefont {L.}~\bibnamefont {Li}}, \bibinfo {author} {\bibfnamefont {N.-L.}\ \bibnamefont {Liu}}, \bibinfo {author} {\bibfnamefont {F.}~\bibnamefont {Xu}}, \bibinfo {author} {\bibfnamefont {Y.-A.}\ \bibnamefont {Chen}},\ and\ \bibinfo {author} {\bibfnamefont {J.-W.}\ \bibnamefont {Pan}},\ }\bibfield  {title} {\bibinfo {title} {Distributed quantum phase estimation with entangled photons},\ }\href {https://www.nature.com/articles/s41566-020-00718-2} {\bibfield  {journal} {\bibinfo  {journal} {Nature photonics}\ }\textbf {\bibinfo {volume} {15}},\ \bibinfo {pages} {137} (\bibinfo {year}
  {2021})}\BibitemShut {NoStop}%
\bibitem [{\citenamefont {Kim}\ \emph {et~al.}(2024)\citenamefont {Kim}, \citenamefont {Hong}, \citenamefont {Kim}, \citenamefont {Kim}, \citenamefont {Lee}, \citenamefont {Pooser}, \citenamefont {Oh}, \citenamefont {Lee}, \citenamefont {Lee},\ and\ \citenamefont {Lim}}]{Kim2024}%
  \BibitemOpen
  \bibfield  {author} {\bibinfo {author} {\bibfnamefont {D.-H.}\ \bibnamefont {Kim}}, \bibinfo {author} {\bibfnamefont {S.}~\bibnamefont {Hong}}, \bibinfo {author} {\bibfnamefont {Y.-S.}\ \bibnamefont {Kim}}, \bibinfo {author} {\bibfnamefont {Y.}~\bibnamefont {Kim}}, \bibinfo {author} {\bibfnamefont {S.-W.}\ \bibnamefont {Lee}}, \bibinfo {author} {\bibfnamefont {R.~C.}\ \bibnamefont {Pooser}}, \bibinfo {author} {\bibfnamefont {K.}~\bibnamefont {Oh}}, \bibinfo {author} {\bibfnamefont {S.-Y.}\ \bibnamefont {Lee}}, \bibinfo {author} {\bibfnamefont {C.}~\bibnamefont {Lee}},\ and\ \bibinfo {author} {\bibfnamefont {H.-T.}\ \bibnamefont {Lim}},\ }\bibfield  {title} {\bibinfo {title} {Distributed quantum sensing of multiple phases with fewer photons},\ }\href {https://www.nature.com/articles/s41467-023-44204-z} {\bibfield  {journal} {\bibinfo  {journal} {Nature communications}\ }\textbf {\bibinfo {volume} {15}},\ \bibinfo {pages} {266} (\bibinfo {year} {2024})}\BibitemShut {NoStop}%
\bibitem [{\citenamefont {Main}\ \emph {et~al.}(2025)\citenamefont {Main}, \citenamefont {Drmota}, \citenamefont {Nadlinger}, \citenamefont {Ainley}, \citenamefont {Agrawal}, \citenamefont {Nichol}, \citenamefont {Srinivas}, \citenamefont {Araneda},\ and\ \citenamefont {Lucas}}]{oxforddc}%
  \BibitemOpen
  \bibfield  {author} {\bibinfo {author} {\bibfnamefont {D.}~\bibnamefont {Main}}, \bibinfo {author} {\bibfnamefont {P.}~\bibnamefont {Drmota}}, \bibinfo {author} {\bibfnamefont {D.~P.}\ \bibnamefont {Nadlinger}}, \bibinfo {author} {\bibfnamefont {E.~M.}\ \bibnamefont {Ainley}}, \bibinfo {author} {\bibfnamefont {A.}~\bibnamefont {Agrawal}}, \bibinfo {author} {\bibfnamefont {B.~C.}\ \bibnamefont {Nichol}}, \bibinfo {author} {\bibfnamefont {R.}~\bibnamefont {Srinivas}}, \bibinfo {author} {\bibfnamefont {G.}~\bibnamefont {Araneda}},\ and\ \bibinfo {author} {\bibfnamefont {D.~M.}\ \bibnamefont {Lucas}},\ }\bibfield  {title} {\bibinfo {title} {Distributed quantum computing across an optical network link},\ }\href {https://www.nature.com/articles/s41586-024-08404-x} {\bibfield  {journal} {\bibinfo  {journal} {Nature (London)}\ }\textbf {\bibinfo {volume} {638}},\ \bibinfo {pages} {383} (\bibinfo {year} {2025})}\BibitemShut {NoStop}%
\bibitem [{\citenamefont {Craddock}\ \emph {et~al.}(2024)\citenamefont {Craddock}, \citenamefont {Lazenby}, \citenamefont {Portmann}, \citenamefont {Sekelsky}, \citenamefont {Flament},\ and\ \citenamefont {Namazi}}]{Craddock2024}%
  \BibitemOpen
  \bibfield  {author} {\bibinfo {author} {\bibfnamefont {A.~N.}\ \bibnamefont {Craddock}}, \bibinfo {author} {\bibfnamefont {A.}~\bibnamefont {Lazenby}}, \bibinfo {author} {\bibfnamefont {G.~B.}\ \bibnamefont {Portmann}}, \bibinfo {author} {\bibfnamefont {R.}~\bibnamefont {Sekelsky}}, \bibinfo {author} {\bibfnamefont {M.}~\bibnamefont {Flament}},\ and\ \bibinfo {author} {\bibfnamefont {M.}~\bibnamefont {Namazi}},\ }\bibfield  {title} {\bibinfo {title} {Automated distribution of polarization-entangled photons using deployed new york city fibers},\ }\href {https://doi.org/10.1103/PRXQuantum.5.030330} {\bibfield  {journal} {\bibinfo  {journal} {PRX Quantum}\ }\textbf {\bibinfo {volume} {5}},\ \bibinfo {pages} {030330} (\bibinfo {year} {2024})}\BibitemShut {NoStop}%
\bibitem [{\citenamefont {Neumann}\ \emph {et~al.}()\citenamefont {Neumann}, \citenamefont {Buchner}, \citenamefont {Bulla}, \citenamefont {Bohmann},\ and\ \citenamefont {Ursin}}]{Neumann2022}%
  \BibitemOpen
  \bibfield  {author} {\bibinfo {author} {\bibfnamefont {S.~P.}\ \bibnamefont {Neumann}}, \bibinfo {author} {\bibfnamefont {A.}~\bibnamefont {Buchner}}, \bibinfo {author} {\bibfnamefont {L.}~\bibnamefont {Bulla}}, \bibinfo {author} {\bibfnamefont {M.}~\bibnamefont {Bohmann}},\ and\ \bibinfo {author} {\bibfnamefont {R.}~\bibnamefont {Ursin}},\ }\bibfield  {title} {\bibinfo {title} {Continuous entanglement distribution over a transnational 248 km fiber link},\ }\href {https://doi.org/https://www.nature.com/articles/s41467-022-33919-0} {\bibfield  {journal} {\bibinfo  {journal} {Nature communications}\ }\textbf {\bibinfo {volume} {13}},\ \bibinfo {pages} {6134}}\BibitemShut {NoStop}%
\bibitem [{\citenamefont {Sharypov}\ and\ \citenamefont {Wilson-Gordon}(2011)}]{Sharypov2011}%
  \BibitemOpen
  \bibfield  {author} {\bibinfo {author} {\bibfnamefont {A.~V.}\ \bibnamefont {Sharypov}}\ and\ \bibinfo {author} {\bibfnamefont {A.~D.}\ \bibnamefont {Wilson-Gordon}},\ }\bibfield  {title} {\bibinfo {title} {Narrowband-biphoton generation due to long-lived coherent population oscillations},\ }\href {https://doi.org/10.1103/PhysRevA.84.033845} {\bibfield  {journal} {\bibinfo  {journal} {Phys. Rev. A}\ }\textbf {\bibinfo {volume} {84}},\ \bibinfo {pages} {033845} (\bibinfo {year} {2011})}\BibitemShut {NoStop}%
\bibitem [{\citenamefont {Zhao}\ \emph {et~al.}(2014)\citenamefont {Zhao}, \citenamefont {Guo}, \citenamefont {Liu}, \citenamefont {Sun}, \citenamefont {Loy},\ and\ \citenamefont {Du}}]{Zhao2014}%
  \BibitemOpen
  \bibfield  {author} {\bibinfo {author} {\bibfnamefont {L.}~\bibnamefont {Zhao}}, \bibinfo {author} {\bibfnamefont {X.}~\bibnamefont {Guo}}, \bibinfo {author} {\bibfnamefont {C.}~\bibnamefont {Liu}}, \bibinfo {author} {\bibfnamefont {Y.}~\bibnamefont {Sun}}, \bibinfo {author} {\bibfnamefont {M.~M.~T.}\ \bibnamefont {Loy}},\ and\ \bibinfo {author} {\bibfnamefont {S.}~\bibnamefont {Du}},\ }\bibfield  {title} {\bibinfo {title} {Photon pairs with coherence time exceeding 1 $\mu$s},\ }\href {https://doi.org/10.1364/OPTICA.1.000084} {\bibfield  {journal} {\bibinfo  {journal} {Optica}\ }\textbf {\bibinfo {volume} {1}},\ \bibinfo {pages} {84} (\bibinfo {year} {2014})}\BibitemShut {NoStop}%
\bibitem [{\citenamefont {Wang}\ \emph {et~al.}(2022)\citenamefont {Wang}, \citenamefont {Li}, \citenamefont {Chang}, \citenamefont {Lin}, \citenamefont {Li}, \citenamefont {Hsiao}, \citenamefont {Chen}, \citenamefont {Lai}, \citenamefont {Chen}, \citenamefont {Chen}, \citenamefont {Chuu},\ and\ \citenamefont {Yu}}]{Wang2022}%
  \BibitemOpen
  \bibfield  {author} {\bibinfo {author} {\bibfnamefont {Y.-S.}\ \bibnamefont {Wang}}, \bibinfo {author} {\bibfnamefont {K.-B.}\ \bibnamefont {Li}}, \bibinfo {author} {\bibfnamefont {C.-F.}\ \bibnamefont {Chang}}, \bibinfo {author} {\bibfnamefont {T.-W.}\ \bibnamefont {Lin}}, \bibinfo {author} {\bibfnamefont {J.-Q.}\ \bibnamefont {Li}}, \bibinfo {author} {\bibfnamefont {S.-S.}\ \bibnamefont {Hsiao}}, \bibinfo {author} {\bibfnamefont {J.-M.}\ \bibnamefont {Chen}}, \bibinfo {author} {\bibfnamefont {Y.-H.}\ \bibnamefont {Lai}}, \bibinfo {author} {\bibfnamefont {Y.-C.}\ \bibnamefont {Chen}}, \bibinfo {author} {\bibfnamefont {Y.-F.}\ \bibnamefont {Chen}}, \bibinfo {author} {\bibfnamefont {C.-S.}\ \bibnamefont {Chuu}},\ and\ \bibinfo {author} {\bibfnamefont {I.~A.}\ \bibnamefont {Yu}},\ }\bibfield  {title} {\bibinfo {title} {Temporally ultralong biphotons with a linewidth of 50 khz},\ }\href {https://doi.org/10.1063/5.0102393} {\bibfield  {journal} {\bibinfo  {journal} {APL Photonics}\ }\textbf {\bibinfo {volume}
  {7}},\ \bibinfo {pages} {126102} (\bibinfo {year} {2022})}\BibitemShut {NoStop}%
\bibitem [{\citenamefont {Li}\ \emph {et~al.}(2023)\citenamefont {Li}, \citenamefont {Li}, \citenamefont {Cao}, \citenamefont {Yin},\ and\ \citenamefont {Peng}}]{Li2023}%
  \BibitemOpen
  \bibfield  {author} {\bibinfo {author} {\bibfnamefont {B.}~\bibnamefont {Li}}, \bibinfo {author} {\bibfnamefont {Y.-H.}\ \bibnamefont {Li}}, \bibinfo {author} {\bibfnamefont {Y.}~\bibnamefont {Cao}}, \bibinfo {author} {\bibfnamefont {J.}~\bibnamefont {Yin}},\ and\ \bibinfo {author} {\bibfnamefont {C.-Z.}\ \bibnamefont {Peng}},\ }\bibfield  {title} {\bibinfo {title} {Pure-state photon-pair source with a long coherence time for large-scale quantum information processing},\ }\href {https://doi.org/10.1103/PhysRevApplied.19.064083} {\bibfield  {journal} {\bibinfo  {journal} {Phys. Rev. Appl.}\ }\textbf {\bibinfo {volume} {19}},\ \bibinfo {pages} {064083} (\bibinfo {year} {2023})}\BibitemShut {NoStop}%
\bibitem [{\citenamefont {Burnham}\ and\ \citenamefont {Weinberg}(1970)}]{burnham1970observation}%
  \BibitemOpen
  \bibfield  {author} {\bibinfo {author} {\bibfnamefont {D.~C.}\ \bibnamefont {Burnham}}\ and\ \bibinfo {author} {\bibfnamefont {D.~L.}\ \bibnamefont {Weinberg}},\ }\bibfield  {title} {\bibinfo {title} {Observation of simultaneity in parametric production of optical photon pairs},\ }\href {https://doi.org/10.1103/PhysRevLett.25.84} {\bibfield  {journal} {\bibinfo  {journal} {Phys. Rev. Lett.}\ }\textbf {\bibinfo {volume} {25}},\ \bibinfo {pages} {84} (\bibinfo {year} {1970})}\BibitemShut {NoStop}%
\bibitem [{\citenamefont {Harris}\ \emph {et~al.}(1967)\citenamefont {Harris}, \citenamefont {Oshman},\ and\ \citenamefont {Byer}}]{harris1967observation}%
  \BibitemOpen
  \bibfield  {author} {\bibinfo {author} {\bibfnamefont {S.~E.}\ \bibnamefont {Harris}}, \bibinfo {author} {\bibfnamefont {M.~K.}\ \bibnamefont {Oshman}},\ and\ \bibinfo {author} {\bibfnamefont {R.~L.}\ \bibnamefont {Byer}},\ }\bibfield  {title} {\bibinfo {title} {Observation of tunable optical parametric fluorescence},\ }\href {https://doi.org/10.1103/PhysRevLett.18.732} {\bibfield  {journal} {\bibinfo  {journal} {Phys. Rev. Lett.}\ }\textbf {\bibinfo {volume} {18}},\ \bibinfo {pages} {732} (\bibinfo {year} {1967})}\BibitemShut {NoStop}%
\bibitem [{\citenamefont {Shih}\ and\ \citenamefont {Alley}(1988)}]{shih1988new}%
  \BibitemOpen
  \bibfield  {author} {\bibinfo {author} {\bibfnamefont {Y.~H.}\ \bibnamefont {Shih}}\ and\ \bibinfo {author} {\bibfnamefont {C.~O.}\ \bibnamefont {Alley}},\ }\bibfield  {title} {\bibinfo {title} {New type of einstein-podolsky-rosen-bohm experiment using pairs of light quanta produced by optical parametric down conversion},\ }\href {https://doi.org/10.1103/PhysRevLett.61.2921} {\bibfield  {journal} {\bibinfo  {journal} {Phys. Rev. Lett.}\ }\textbf {\bibinfo {volume} {61}},\ \bibinfo {pages} {2921} (\bibinfo {year} {1988})}\BibitemShut {NoStop}%
\bibitem [{\citenamefont {Kwiat}\ \emph {et~al.}(1995)\citenamefont {Kwiat}, \citenamefont {Mattle}, \citenamefont {Weinfurter}, \citenamefont {Zeilinger}, \citenamefont {Sergienko},\ and\ \citenamefont {Shih}}]{kwiat1995new}%
  \BibitemOpen
  \bibfield  {author} {\bibinfo {author} {\bibfnamefont {P.~G.}\ \bibnamefont {Kwiat}}, \bibinfo {author} {\bibfnamefont {K.}~\bibnamefont {Mattle}}, \bibinfo {author} {\bibfnamefont {H.}~\bibnamefont {Weinfurter}}, \bibinfo {author} {\bibfnamefont {A.}~\bibnamefont {Zeilinger}}, \bibinfo {author} {\bibfnamefont {A.~V.}\ \bibnamefont {Sergienko}},\ and\ \bibinfo {author} {\bibfnamefont {Y.}~\bibnamefont {Shih}},\ }\bibfield  {title} {\bibinfo {title} {New high-intensity source of polarization-entangled photon pairs},\ }\href {https://doi.org/10.1103/PhysRevLett.75.4337} {\bibfield  {journal} {\bibinfo  {journal} {Phys. Rev. Lett.}\ }\textbf {\bibinfo {volume} {75}},\ \bibinfo {pages} {4337} (\bibinfo {year} {1995})}\BibitemShut {NoStop}%
\bibitem [{\citenamefont {Hong}\ \emph {et~al.}(1987)\citenamefont {Hong}, \citenamefont {Ou},\ and\ \citenamefont {Mandel}}]{hong1987measurement}%
  \BibitemOpen
  \bibfield  {author} {\bibinfo {author} {\bibfnamefont {C.~K.}\ \bibnamefont {Hong}}, \bibinfo {author} {\bibfnamefont {Z.~Y.}\ \bibnamefont {Ou}},\ and\ \bibinfo {author} {\bibfnamefont {L.}~\bibnamefont {Mandel}},\ }\bibfield  {title} {\bibinfo {title} {Measurement of subpicosecond time intervals between two photons by interference},\ }\href {https://doi.org/10.1103/PhysRevLett.59.2044} {\bibfield  {journal} {\bibinfo  {journal} {Phys. Rev. Lett.}\ }\textbf {\bibinfo {volume} {59}},\ \bibinfo {pages} {2044} (\bibinfo {year} {1987})}\BibitemShut {NoStop}%
\bibitem [{\citenamefont {Liu}\ \emph {et~al.}(2020)\citenamefont {Liu}, \citenamefont {Liu}, \citenamefont {Yu},\ and\ \citenamefont {Zhang}}]{liu2020sub}%
  \BibitemOpen
  \bibfield  {author} {\bibinfo {author} {\bibfnamefont {J.}~\bibnamefont {Liu}}, \bibinfo {author} {\bibfnamefont {J.}~\bibnamefont {Liu}}, \bibinfo {author} {\bibfnamefont {P.}~\bibnamefont {Yu}},\ and\ \bibinfo {author} {\bibfnamefont {G.}~\bibnamefont {Zhang}},\ }\bibfield  {title} {\bibinfo {title} {Sub-megahertz narrow-band photon pairs at 606 nm for solid-state quantum memories},\ }\href {https://doi.org/10.1063/5.0006021} {\bibfield  {journal} {\bibinfo  {journal} {APL Photonics}\ }\textbf {\bibinfo {volume} {5}},\ \bibinfo {pages} {066105} (\bibinfo {year} {2020})}\BibitemShut {NoStop}%
\bibitem [{\citenamefont {Du}\ \emph {et~al.}(2008{\natexlab{a}})\citenamefont {Du}, \citenamefont {Kolchin}, \citenamefont {Belthangady}, \citenamefont {Yin},\ and\ \citenamefont {Harris}}]{PhysRevLett.100.183603}%
  \BibitemOpen
  \bibfield  {author} {\bibinfo {author} {\bibfnamefont {S.}~\bibnamefont {Du}}, \bibinfo {author} {\bibfnamefont {P.}~\bibnamefont {Kolchin}}, \bibinfo {author} {\bibfnamefont {C.}~\bibnamefont {Belthangady}}, \bibinfo {author} {\bibfnamefont {G.~Y.}\ \bibnamefont {Yin}},\ and\ \bibinfo {author} {\bibfnamefont {S.~E.}\ \bibnamefont {Harris}},\ }\bibfield  {title} {\bibinfo {title} {Subnatural linewidth biphotons with controllable temporal length},\ }\href {https://doi.org/10.1103/PhysRevLett.100.183603} {\bibfield  {journal} {\bibinfo  {journal} {Phys. Rev. Lett.}\ }\textbf {\bibinfo {volume} {100}},\ \bibinfo {pages} {183603} (\bibinfo {year} {2008}{\natexlab{a}})}\BibitemShut {NoStop}%
\bibitem [{\citenamefont {Du}\ \emph {et~al.}(2008{\natexlab{b}})\citenamefont {Du}, \citenamefont {Wen},\ and\ \citenamefont {Rubin}}]{Du2008}%
  \BibitemOpen
  \bibfield  {author} {\bibinfo {author} {\bibfnamefont {S.}~\bibnamefont {Du}}, \bibinfo {author} {\bibfnamefont {J.}~\bibnamefont {Wen}},\ and\ \bibinfo {author} {\bibfnamefont {M.~H.}\ \bibnamefont {Rubin}},\ }\bibfield  {title} {\bibinfo {title} {Narrowband biphoton generation near atomic resonance},\ }\href {https://doi.org/10.1364/JOSAB.25.000C98} {\bibfield  {journal} {\bibinfo  {journal} {J. Opt. Soc. Am. B}\ }\textbf {\bibinfo {volume} {25}},\ \bibinfo {pages} {C98} (\bibinfo {year} {2008}{\natexlab{b}})}\BibitemShut {NoStop}%
\bibitem [{\citenamefont {Boller}\ \emph {et~al.}(1991)\citenamefont {Boller}, \citenamefont {Imamo\ifmmode~\breve{g}\else \u{g}\fi{}lu},\ and\ \citenamefont {Harris}}]{boller1991observation}%
  \BibitemOpen
  \bibfield  {author} {\bibinfo {author} {\bibfnamefont {K.-J.}\ \bibnamefont {Boller}}, \bibinfo {author} {\bibfnamefont {A.}~\bibnamefont {Imamo\ifmmode~\breve{g}\else \u{g}\fi{}lu}},\ and\ \bibinfo {author} {\bibfnamefont {S.~E.}\ \bibnamefont {Harris}},\ }\bibfield  {title} {\bibinfo {title} {Observation of electromagnetically induced transparency},\ }\href {https://doi.org/10.1103/PhysRevLett.66.2593} {\bibfield  {journal} {\bibinfo  {journal} {Phys. Rev. Lett.}\ }\textbf {\bibinfo {volume} {66}},\ \bibinfo {pages} {2593} (\bibinfo {year} {1991})}\BibitemShut {NoStop}%
\bibitem [{\citenamefont {Harris}(1997)}]{harris1997electromagnetically}%
  \BibitemOpen
  \bibfield  {author} {\bibinfo {author} {\bibfnamefont {S.~E.}\ \bibnamefont {Harris}},\ }\bibfield  {title} {\bibinfo {title} {Electromagnetically induced transparency},\ }\href {https://doi.org/10.1063/1.881806} {\bibfield  {journal} {\bibinfo  {journal} {Physics Today}\ }\textbf {\bibinfo {volume} {50}},\ \bibinfo {pages} {36} (\bibinfo {year} {1997})}\BibitemShut {NoStop}%
\bibitem [{\citenamefont {Lukin}\ and\ \citenamefont {Imamo{\u{g}}lu}(2001)}]{lukin2001controlling}%
  \BibitemOpen
  \bibfield  {author} {\bibinfo {author} {\bibfnamefont {M.}~\bibnamefont {Lukin}}\ and\ \bibinfo {author} {\bibfnamefont {A.}~\bibnamefont {Imamo{\u{g}}lu}},\ }\bibfield  {title} {\bibinfo {title} {Controlling photons using electromagnetically induced transparency},\ }\href {https://doi.org/10.1038/35095000} {\bibfield  {journal} {\bibinfo  {journal} {Nature}\ }\textbf {\bibinfo {volume} {413}},\ \bibinfo {pages} {273} (\bibinfo {year} {2001})}\BibitemShut {NoStop}%
\bibitem [{\citenamefont {Lai}\ \emph {et~al.}(2024)\citenamefont {Lai}, \citenamefont {Li}, \citenamefont {Zanders}, \citenamefont {Mei},\ and\ \citenamefont {Du}}]{PRLsymbiph}%
  \BibitemOpen
  \bibfield  {author} {\bibinfo {author} {\bibfnamefont {X.}~\bibnamefont {Lai}}, \bibinfo {author} {\bibfnamefont {C.}~\bibnamefont {Li}}, \bibinfo {author} {\bibfnamefont {A.}~\bibnamefont {Zanders}}, \bibinfo {author} {\bibfnamefont {Y.}~\bibnamefont {Mei}},\ and\ \bibinfo {author} {\bibfnamefont {S.}~\bibnamefont {Du}},\ }\bibfield  {title} {\bibinfo {title} {Symmetry protected two-photon coherence time},\ }\href {https://doi.org/10.1103/PhysRevLett.133.033601} {\bibfield  {journal} {\bibinfo  {journal} {Phys. Rev. Lett.}\ }\textbf {\bibinfo {volume} {133}},\ \bibinfo {pages} {033601} (\bibinfo {year} {2024})}\BibitemShut {NoStop}%
\bibitem [{\citenamefont {Jiang}\ \emph {et~al.}(2023)\citenamefont {Jiang}, \citenamefont {Mei},\ and\ \citenamefont {Du}}]{Jiang2023}%
  \BibitemOpen
  \bibfield  {author} {\bibinfo {author} {\bibfnamefont {Y.}~\bibnamefont {Jiang}}, \bibinfo {author} {\bibfnamefont {Y.}~\bibnamefont {Mei}},\ and\ \bibinfo {author} {\bibfnamefont {S.}~\bibnamefont {Du}},\ }\bibfield  {title} {\bibinfo {title} {Quantum langevin theory for two coupled phase-conjugated electromagnetic waves},\ }\href {https://doi.org/10.1103/PhysRevA.107.053703} {\bibfield  {journal} {\bibinfo  {journal} {Phys. Rev. A}\ }\textbf {\bibinfo {volume} {107}},\ \bibinfo {pages} {053703} (\bibinfo {year} {2023})}\BibitemShut {NoStop}%
\bibitem [{\citenamefont {Bender}\ and\ \citenamefont {Boettcher}(1998)}]{PhysRevLett.80.5243}%
  \BibitemOpen
  \bibfield  {author} {\bibinfo {author} {\bibfnamefont {C.~M.}\ \bibnamefont {Bender}}\ and\ \bibinfo {author} {\bibfnamefont {S.}~\bibnamefont {Boettcher}},\ }\bibfield  {title} {\bibinfo {title} {Real spectra in non-hermitian hamiltonians having pt symmetry},\ }\href {https://doi.org/10.1103/PhysRevLett.80.5243} {\bibfield  {journal} {\bibinfo  {journal} {Phys. Rev. Lett.}\ }\textbf {\bibinfo {volume} {80}},\ \bibinfo {pages} {5243} (\bibinfo {year} {1998})}\BibitemShut {NoStop}%
\bibitem [{\citenamefont {Miri}\ and\ \citenamefont {Al{\`{u}}}(2019)}]{Miri_2019}%
  \BibitemOpen
  \bibfield  {author} {\bibinfo {author} {\bibfnamefont {M.-A.}\ \bibnamefont {Miri}}\ and\ \bibinfo {author} {\bibfnamefont {A.}~\bibnamefont {Al{\`{u}}}},\ }\bibfield  {title} {\bibinfo {title} {Exceptional points in optics and photonics},\ }\href {https://doi.org/10.1126/science.aar7709} {\bibfield  {journal} {\bibinfo  {journal} {Science}\ }\textbf {\bibinfo {volume} {363}},\ \bibinfo {pages} {eaar7709} (\bibinfo {year} {2019})}\BibitemShut {NoStop}%
\bibitem [{\citenamefont {Rüter}\ \emph {et~al.}(2010)\citenamefont {Rüter}, \citenamefont {Makris}, \citenamefont {El-Ganainy}, \citenamefont {Christodoulides}, \citenamefont {Segev},\ and\ \citenamefont {Kip}}]{Ruter2010}%
  \BibitemOpen
  \bibfield  {author} {\bibinfo {author} {\bibfnamefont {C.~E.}\ \bibnamefont {Rüter}}, \bibinfo {author} {\bibfnamefont {K.~G.}\ \bibnamefont {Makris}}, \bibinfo {author} {\bibfnamefont {R.}~\bibnamefont {El-Ganainy}}, \bibinfo {author} {\bibfnamefont {D.~N.}\ \bibnamefont {Christodoulides}}, \bibinfo {author} {\bibfnamefont {M.}~\bibnamefont {Segev}},\ and\ \bibinfo {author} {\bibfnamefont {D.}~\bibnamefont {Kip}},\ }\bibfield  {title} {\bibinfo {title} {Observation of parity–time symmetry in optics},\ }\href {https://doi.org/10.1038/nphys1515} {\bibfield  {journal} {\bibinfo  {journal} {Nature Physics}\ }\textbf {\bibinfo {volume} {6}},\ \bibinfo {pages} {192} (\bibinfo {year} {2010})}\BibitemShut {NoStop}%
\bibitem [{\citenamefont {Lupu}\ \emph {et~al.}(2013)\citenamefont {Lupu}, \citenamefont {Benisty},\ and\ \citenamefont {Degiron}}]{Lupu2013}%
  \BibitemOpen
  \bibfield  {author} {\bibinfo {author} {\bibfnamefont {A.}~\bibnamefont {Lupu}}, \bibinfo {author} {\bibfnamefont {H.}~\bibnamefont {Benisty}},\ and\ \bibinfo {author} {\bibfnamefont {A.}~\bibnamefont {Degiron}},\ }\bibfield  {title} {\bibinfo {title} {Switching using pt symmetry in plasmonic systems: positive role of the losses},\ }\href {https://doi.org/10.1364/OE.21.021651} {\bibfield  {journal} {\bibinfo  {journal} {Opt. Express}\ }\textbf {\bibinfo {volume} {21}},\ \bibinfo {pages} {21651} (\bibinfo {year} {2013})}\BibitemShut {NoStop}%
\bibitem [{\citenamefont {Liu}\ \emph {et~al.}(2016)\citenamefont {Liu}, \citenamefont {Zhang}, \citenamefont {Ozdemir}, \citenamefont {Peng}, \citenamefont {Jing}, \citenamefont {Lü}, \citenamefont {Li}, \citenamefont {Yang}, \citenamefont {Nori},\ and\ \citenamefont {Liu}}]{Liu2016}%
  \BibitemOpen
  \bibfield  {author} {\bibinfo {author} {\bibfnamefont {Z.-P.}\ \bibnamefont {Liu}}, \bibinfo {author} {\bibfnamefont {J.}~\bibnamefont {Zhang}}, \bibinfo {author} {\bibfnamefont {S.~K.}\ \bibnamefont {Ozdemir}}, \bibinfo {author} {\bibfnamefont {B.}~\bibnamefont {Peng}}, \bibinfo {author} {\bibfnamefont {H.}~\bibnamefont {Jing}}, \bibinfo {author} {\bibfnamefont {X.-Y.}\ \bibnamefont {Lü}}, \bibinfo {author} {\bibfnamefont {C.-W.}\ \bibnamefont {Li}}, \bibinfo {author} {\bibfnamefont {L.}~\bibnamefont {Yang}}, \bibinfo {author} {\bibfnamefont {F.}~\bibnamefont {Nori}},\ and\ \bibinfo {author} {\bibfnamefont {Y.-x.}\ \bibnamefont {Liu}},\ }\bibfield  {title} {\bibinfo {title} {Metrology with pt-symmetric cavities: Enhanced sensitivity near the pt-phase transition},\ }\href {https://doi.org/10.1103/PhysRevLett.117.110802} {\bibfield  {journal} {\bibinfo  {journal} {Phys. Rev. Lett.}\ }\textbf {\bibinfo {volume} {117}},\ \bibinfo {pages} {110802} (\bibinfo {year} {2016})}\BibitemShut {NoStop}%
\bibitem [{\citenamefont {Glauber}(2006)}]{Glauber2006}%
  \BibitemOpen
  \bibfield  {author} {\bibinfo {author} {\bibfnamefont {R.~J.}\ \bibnamefont {Glauber}},\ }\bibfield  {title} {\bibinfo {title} {Nobel lecture: One hundred years of light quanta},\ }\href {https://doi.org/10.1103/RevModPhys.78.1267} {\bibfield  {journal} {\bibinfo  {journal} {Rev. Mod. Phys.}\ }\textbf {\bibinfo {volume} {78}},\ \bibinfo {pages} {1267} (\bibinfo {year} {2006})}\BibitemShut {NoStop}%
\end{thebibliography}%

\end{document}